\documentclass[12pt,thmsa]{article}
\usepackage{amssymb}
\usepackage{graphicx}

\input{tcilatex}
\begin{document}

\begin{center}
{\large A qualitative perspective on the dynamics of a
single-Cooper-pair box with a phase-damped cavity}

\textbf{M. Abdel-Aty}\footnote{%
E-mail: abdelatyquant@yahoo.co.uk}

{\footnotesize Mathematics Department, Faculty of Science, South
Valley University, 82524 Sohag, Egypt \\ Mathematics Department,
College of Science, Bahrain University, 32038 Kingdom of Bahrain }

~

{\bf  J. Phys. A: Math. Theor. 41 185304  (2008)}

\end{center}

In a recent paper Dajka, et.al., [J. Phys. A \textbf{40}, F879
(2007)] predicted that some composite systems can be entangled
forever even if coupled with a thermal bath. We analyze the
transient entanglement of a single-Cooper-pair box biased by a
classical voltage and irradiated by a quantized field and find the
unusual feature that the phase-damped cavity can lead to a
long-lived entanglement. The results show an asymptotic value of the
idempotency defect (concurrence) which embodies coherence loss
(entanglement survival), independent of the interaction development
by dependent critically on environment.


\section{Introduction}

Josephson junctions are being investigated as a possible route to
scalable quantum computers \cite{lut07}-\cite{vio02}. The present
lack of a current standard based on quantum devices has inspired
several attempts to manipulate single electrons, where the rate of
particle transfer is controlled by an external frequency. One of the
physical realizations of a solid-state qubit is provided by a Cooper
pair box which is a small superconducting island connected to a
large superconducting electrode, a reservoir, through a Josephson
junction \cite{gri07}. Also realizations of superconducting charge
qubits are a promising technology for the realization of quantum
computation on a large scale \cite{coo04}-\cite{ast06}.

In this context, a solid-state system is highly desirable because of its
compactness, scalability and compatibility with existing semiconductor
technology. Even though a Cooper pair box can contains millions of electrons
at any one time, the box exhibits only two quantum charge states, depending
upon whether or not a Cooper pair of electrons has recently tunneled into
the box and various superconducting nanocircuits have been proposed as
quantum bits (qubits) for a quantum computer \cite{nis07,you05}. By gating
the Cooper pairs into the box with an appropriate pulse width, previous
research has shown that a coherent superposition of the two states can
enable quantum computations. In architectures based on Josephson junctions
coupled to resonators, the resonators store single qubit states, transfer
states from one Josephson junction to another, entangle two or more
Josephson junctions, and mediate two-qubit quantum logic. In effect, the
resonators are the quantum computational analog of the classical memory and
bus elements.

The present work is motivated by conjectures and statements
presented in a recent fast track paper \cite{daj07} and experimental
results on Josephson junction and normal metal flux qubits coupled
to the environment \cite{sch01}. We obtain a long-lived entanglement
using a superconducting charge qubit. More precisely, we endeavor to
show the important property of entanglement via idempotency defect
of a single Cooper pair box, due to the presence of a phase-damped
cavity. Despite the complexity of the problem, we obtain a quite
simple exact solution of the master equation that is valid for
arbitrary values of the phase damping. In the framework of the exact
solution of the master equation, we determine the coherence loss and
the degree of entanglement. We perform a systematic analysis in
order to reach an understanding of the Cooper pair dynamics in the
presence of the decoherence. Physically, the effect of phase damping
may be understood to be analogy of the $T_{2}$ spin depolarization
effects observed in nuclear magnetic resonance spectroscopy (for
detailed physical motivation see Ref. \cite{oli00}). Besides
phase-damping-model importance in the description of different
physical situations, it is very instructive since it allows for
obtaining analytical treatments for different entanglement measures
of some classes of states \cite{car07}. Some theoretical discussions
and analysis of special cases of the problem at hand were given in
Refs. \cite{liu05,you03,ste06} and experimental results were
predicted in Ref. \cite{sch01}.

The organization of this paper is as follows: in section 2 we
introduce the model and give the exact solution of the master
equation. In section 3 we employ the analytical results obtained in
section 2 to discuss the idempotency defect and entanglement for
different values of the phase-damped cavity. Finally, we summarize
the results in section 4.

\begin{figure}[tbph]
\begin{center}
\includegraphics[width=10cm,height=8cm]{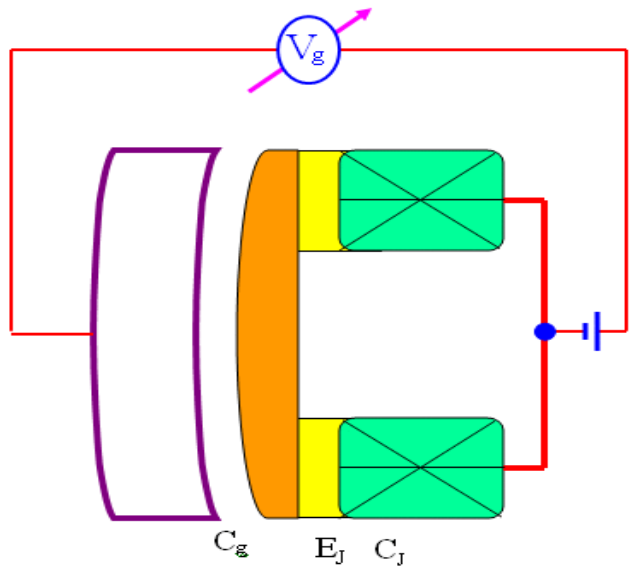}
\end{center}
\caption{Schematic picture of the Cooper-pair box which is driven by an
applied voltage $V_{g}$ through the gate capacitance $C_{g}$. Black bars
denote Cooper pair box. The two Josephson junctions have capacitance $C_{J}$
and Josephson energy $E_{J}$. The driving microwave field is generated using
the electrical voltage acting on the charge qubit via the gate capacitance
\protect\cite{vio02}. }
\label{sch}
\end{figure}

\section{The model}

Several schemes have been proposed for implementing quantum computer
hardware in solid state quantum electronics \cite{mak99}. These
schemes use electric charge, magnetic flux, superconducting phase,
electron spin, or nuclear spin as the information bearing degree of
freedom.

We start our analysis by presenting a brief discussion and a few
physical principles of the Cooper pair box system. We consider a
superconducting box with a low-capacitance Josephson junction with
the capacitance $C_{J}$ and Josephson energy $E_{J}$, biased by a
classical voltage source $V_{g}$ through a gate capacitance $C_{g}$
and placed inside a single-mode microwave cavity. In particular, the
schematic picture of this single-qubit structure may be modeled as
shown in figure 1. The total Hamiltonian of the system can be
written as \cite{mig01}
\begin{equation}
\hat{H}=\frac{1}{2}(Q-C_{g}V_{g}-C_{g}V)^{2}(C_{g}+C_{J})^{-1}-E_{J}\cos
\phi +\hbar \omega (\psi ^{\dagger }\psi +\frac{1}{2}),  \nonumber  \label{5}
\end{equation}%
where $Q=2Ne$ is the charge on the island ($e$ is the electron
charge and $N$ is the number of Cooper-pairs) and $\phi $ is the
phase difference across the junction. The radiation field is to
produce an alternating electric field of the same frequency across
the junction, and $V$ $\ $is the effective voltage difference
produced by the microwave across the junction. We assume that the
dimension of the device is much smaller than the wavelength of the
applied quantized microwave (which is a realistic assumption), so
the spatial variation in the electric field is negligible. We also
assume that the field is linearly polarized, and is taken
perpendicular to the plane of electrodes, then $V$ $\ $can be
written as \cite{you03,zha02} $V=i\hslash \omega \left(
\hat{\psi}-\hat{\psi}^{\dagger }\right) /(2C_{F}),$ where
$\hat{\psi}^{\dagger }$ and $\hat{\psi}$ are the
creation and annihilation operators of the microwave field with frequency $%
\omega $. We denote the capacitance parameter by $C_{F}$, which depends on
the thickness of the junction, the relative dielectric constant of the thin
insulating barrier, and the dimension of the cavity.

We consider the case where the charging energy with scale $E_{c}=\frac{1}{2}%
e^{2}\left( C_{g}+C_{J}\right),$ which dominates over the Josephson
coupling energy $E_{_{J}}$ and concentrates on the value
$V_{g}=e/C_{g}$, so that only the low-energy charge states $N=0$ and
$N=1$ are relevant. In this case the Hamiltonian, in the basis of
the charge states $\left\vert 0\right\rangle $ and $\left\vert
1\right\rangle $, reduces to a two-state form. In a spin-$1/2$
language \cite{kre00}
\begin{equation}
\hat{H}=E_{c}\left( 1+e^{-2}C_{J}^{2}V^{2}\right) -\frac{1}{2}E_{J}\sigma
_{x}+2e^{-1}E_{c}C_{J}V\sigma _{z}+\hbar \omega \left( \psi ^{\dagger }\psi +%
\frac{1}{2}\right) ,  \label{ham}
\end{equation}%
where $\sigma _{x}$ and $\sigma _{z}$ are the Pauli matrices in the
pseudo-spin basis. It is to be noted that the charge states are not
the eigenstates of the Hamiltonian (\ref{ham}), so the Hamiltonian
can be diagonalized yielding the following two charge states
subspace $|e\rangle =\frac{1}{\sqrt{2}}(\left\vert 1\right\rangle
-\left\vert 0\right\rangle)$ and $|g\rangle
=\frac{1}{\sqrt{2}}(\left\vert 1\right\rangle +\left\vert
0\right\rangle ).$ Here we employ these eigenstates to represent the
qubit. If we \textrm{consider a weak quantized radiation field
and neglect the term containing }$V^{2}$\textrm{, the }Hamiltonian\textrm{\ (%
\ref{ham}) can be rewritten in the rotating wave approximations as }%
\begin{equation}
\hat{H}=\hbar \omega \psi ^{\dagger }\psi +\frac{1}{2}E_{J}\sigma
_{z}+\left\{ \frac{-i\hbar eC_{J}}{2(C_{g}+C_{J})}\sqrt{\frac{\omega }{%
2\hbar C_{F}}}\psi \sigma _{+}+H.c.\right\} .
\end{equation}%
We consider the interaction with an environment to be as the phase-damping
type. This is a reservoir coupled to the field via the number operator of
the indicating field, so that there is no energy damping, although there is
a phase damping.

In order to obtain the general solution of the master equation for
the density matrix under the phase damping of the cavity field at a
zero temperature bath, we write
\begin{equation}
\frac{d\hat{\rho}(t)}{dt}=\frac{-i}{\hbar }[\hat{H},\hat{\rho}(t)]+\gamma
\left( 2\hat{\psi}^{\dag }\hat{\psi}\hat{\rho}(t)\hat{\psi}^{\dag }\hat{\psi}%
-\hat{\psi}^{\dag }\hat{\psi}\hat{\psi}^{\dag }\hat{\psi}\hat{\rho}(t)-\hat{%
\rho}(t)\hat{\psi}^{\dag }\hat{\psi}\hat{\psi}^{\dag }\hat{\psi}\right) ,
\label{mas}
\end{equation}%
where $\gamma $ is the phase-damping constant. An important feature
of this quantized system is that its steady states, known as dressed
states, are entangled. Switching to an interaction-picture
representation for convenience by defining $\hat{\rho}^{\prime }(t)=\exp (i%
\hat{H}t)\hat{\rho}(t)\exp (-i\hat{H}t)$, exact solution of equation
(4) can be obtained in the dressed-states representations
\cite{pur87}. Consequently, equation (\ref{mas}) can be written as
\begin{equation}
\frac{d\hat{\rho}^{\prime }(t)}{dt}=\gamma e^{i\hat{H}t}\left( 2\hat{\psi}%
^{\dag }\hat{\psi}\hat{\rho}(t)\hat{\psi}^{\dag }\hat{\psi}-\hat{\psi}^{\dag
}\hat{\psi}\hat{\psi}^{\dag }\hat{\psi}\hat{\rho}(t)-\hat{\rho}(t)\hat{\psi}%
^{\dag }\hat{\psi}\hat{\psi}^{\dag }\hat{\psi}\right) e^{-i\hat{H}t}.
\label{mas2}
\end{equation}%
Next, we write the field operators $\hat{\psi}$ and
$\hat{\psi}^{\dag }\hat{\psi}$ in terms of the dressed states basis
and get the initial state of the system expressed in the product
density matrix forms. In the basis $|n,e\rangle $ and $|n,g\rangle $
states, the field operator $\hat{\psi}$ can be written as
$\hat{\psi}=$ $\sum\limits_{n=0}^{\infty }$ $\sqrt{n}(|n-1,e\rangle
\langle n,e|+|n-1,g\rangle \langle n,g|)$. Neglecting the
oscillating terms of the master equation (\ref{mas2}) in secular
approximation, the density matrix in terms of the dressed states
becomes
\begin{eqnarray}
\frac{d\hat{\rho}^{\prime }(t)}{dt} &=&-\frac{\gamma }{4}\sum\limits_{n=0}^{%
\infty }\left\{ \left( |\phi _{n}^{(+)}\rangle \langle \phi _{n}^{(+)}|\hat{%
\rho}^{\prime }(t)|\phi _{m}^{(+)}\rangle \langle \phi _{m}^{(+)}|\right.
\right.   \nonumber \\
&&+|\phi _{n}^{(-)}\rangle \langle \phi _{n}^{(-)}|\hat{\rho}^{\prime
}(t)|\phi _{m}^{(-)}\rangle \langle \phi _{m}^{(-)}|+|\phi _{n}^{(+)}\rangle
\langle \phi _{n}^{(+)}|\hat{\rho}^{\prime }(t)|\phi _{m}^{(-)}\rangle
\langle \phi _{m}^{(-)}|  \nonumber \\
&&\left. +|\phi _{n}^{(+)}\rangle \langle \phi _{n}^{(+)}|\hat{\rho}^{\prime
}(t)|\phi _{m}^{(-)}\rangle \langle \phi _{m}^{(-)}|\right)
((2n+1)^{2}+1)+|\phi _{n}^{(+)}\rangle \langle \phi _{n}^{(-)}|\hat{\rho}%
^{\prime }(t)  \nonumber \\
&&\times |\phi _{m}^{(-)}\rangle \langle \phi _{m}^{(+)}|e^{2it\mu
_{nm}}+|\phi _{n}^{(-)}\rangle \langle \phi _{n}^{(+)}|\hat{\rho}^{\prime
}(t)|\phi _{m}^{(+)}\rangle \langle \phi _{m}^{(-)}|e^{-2it\mu _{nm}}
\nonumber  \label{mas3} \\
&&+\frac{\gamma }{2}\sum\limits_{n,m=0}^{\infty }\left\{ ((2n+1)(2m+1)(\phi
_{nm}^{(+-)}\hat{\rho}^{\prime }(t)+\hat{\rho}^{\prime }(t)\phi
_{nm}^{(+-)})\right\} ,
\end{eqnarray}%
where $|\phi _{n}^{(\pm )}\rangle $ are two eigenstates of the Hamiltonian
(3) for a lossless cavity, $\phi _{nm}^{(+-)}=|\phi _{n}^{(+)}\rangle
\langle \phi _{n}^{(+)}|+|\phi _{m}^{(+)}\rangle \langle \phi _{m}^{(+)}|$
and $\mu _{nm}=\mu _{n}-\mu _{m}$. The eigenvalues are given by $\pm \mu
_{n},$ where
\begin{equation}
\mu _{n}=\frac{1}{2(C_{J}+C_{g})}\sqrt{\frac{8\delta
^{2}C_{F}(C_{J}+C_{g})^{2}+e^{2}\omega C_{J}^{2}(n+1)}{2C_{F}}}.
\end{equation}%
We denote by $\Delta =E_{J}-\omega $ the detuning between the
Josephson energy and cavity field frequency, ($\delta =\Delta /2)$.
Based on the preparatory work, now we can find an exact solution
under certain conditions of the whole system. With this in mind we
will assume that the initial state is prepared to be a particular
coherent state $\rho ^{f}(0)=|\alpha \rangle \langle \alpha |$ of
the field with the Cooper pair box prepared in the state $\rho
^{J}(0)=|e\rangle \langle e|$. The initial state of the system can
be expressed in the product density matrix form, $\rho (0)=$ $\rho
^{J}(0)\otimes $ $\rho ^{f}(0).$ Consequently, the general solution
to equation (\ref{mas3}) my be written explicitly as
\begin{eqnarray}
\hat{\rho}(t) &=&\sum\limits_{n,m=0}^{\infty }b_{n}b_{m}^{\ast }\exp \left(
\frac{-\gamma }{2}t\right) \exp \left( -\gamma t(n-m)^{2}\right)   \nonumber
\\
&&\times \left\{ \exp (-i\beta _{12})\left( \cos \left( \mu _{nm}(t)\right)
+\cos \left( \mu _{nm}^{\prime }(t)\right) \right) \left\vert
n,e\right\rangle \left\langle m,e\right\vert \frac{{}}{{}}\right.   \nonumber
\\
&&-\frac{i}{2}\exp (-i\beta _{12})\sin \left( \mu _{nm}(t)\right) \left\vert
n,e\right\rangle \left\langle m+1,g\right\vert   \nonumber \\
&&+\frac{i}{2}\exp (i\beta _{12})\sin \left( \mu _{nm}(t)\right) \left\vert
n+1,g\right\rangle \left\langle m,e\right\vert +\exp (-i\beta _{12})
\nonumber \\
&&\left. \frac{{}}{{}}\times \left( \cos \left( \mu _{nm}(t)\right) -\cos
\left( \mu _{nm}^{\prime }(t)\right) \right) \left\vert n+1,g\right\rangle
\left\langle m+1,g\right\vert \right\} ,  \label{exa}
\end{eqnarray}%
where $\mu _{nm}(t)=\mu _{n}(t)-\mu _{m}(t),$ $\mu _{nm}^{\prime }(t)=\mu
_{n}(t)+\mu _{m}(t).$ The probability distribution among Fock states is
Poissonian, $b_{n}=\langle n|\alpha \rangle ,$ with $\overline{n}=|\alpha
|^{2}$ and $\beta _{12}=\beta -\beta ^{\ast },$ where $\beta $ is the phase
of the initial state of the field i.e. $\alpha =|\alpha |e^{i\beta }.$ The
decoherence effect on the dynamical evolution of the present system can be
discussed through the phase-damping constant $\gamma $.

\section{ Coherence loss and entanglement}
In general, due to decoherence, a pure state is apt to change into a
mixed state. However, in many cases of quantum information
processing, one requires a state with high purity and large amount
of entanglement. Therefore, it is necessary to consider the mixture
of the state and its relation with entanglement.

Here we use the idempotency defect, defined by linear entropy, as a
measure of the degree of mixture for a state $\hat{\rho}^{J}(t)$, in
analogy to what is done for the calculation of the entanglement in
terms of von Neumann entropy \cite{san00} which has similar
behavior. In order to analyze what happens to the Cooper pair box,
we trace out the field variables from the state $\hat{\rho}(t)$ and
get the reduced density matrix
$\hat{\rho}^{J}(t)=tr_{f}\hat{\rho}(t)$. The idempotency defect as a
measure of coherence loss can be written as
\begin{equation}
\mathcal{E}_{t}^{(J)}=Tr\left\{ \hat{\rho}_{J}(t)(1-\hat{\rho}_{J}(t))\frac{%
{}}{{}}\right\} ,{\ \ \ \ \ \ \ \ }  \label{lin}
\end{equation}%
where $\mathcal{E}_{t}^{(J)}$ has a zero value for a pure state and $1$ for
a completely mixed state.

\begin{figure}[tbph]
\begin{center}
\includegraphics[width=10cm,height=8cm]{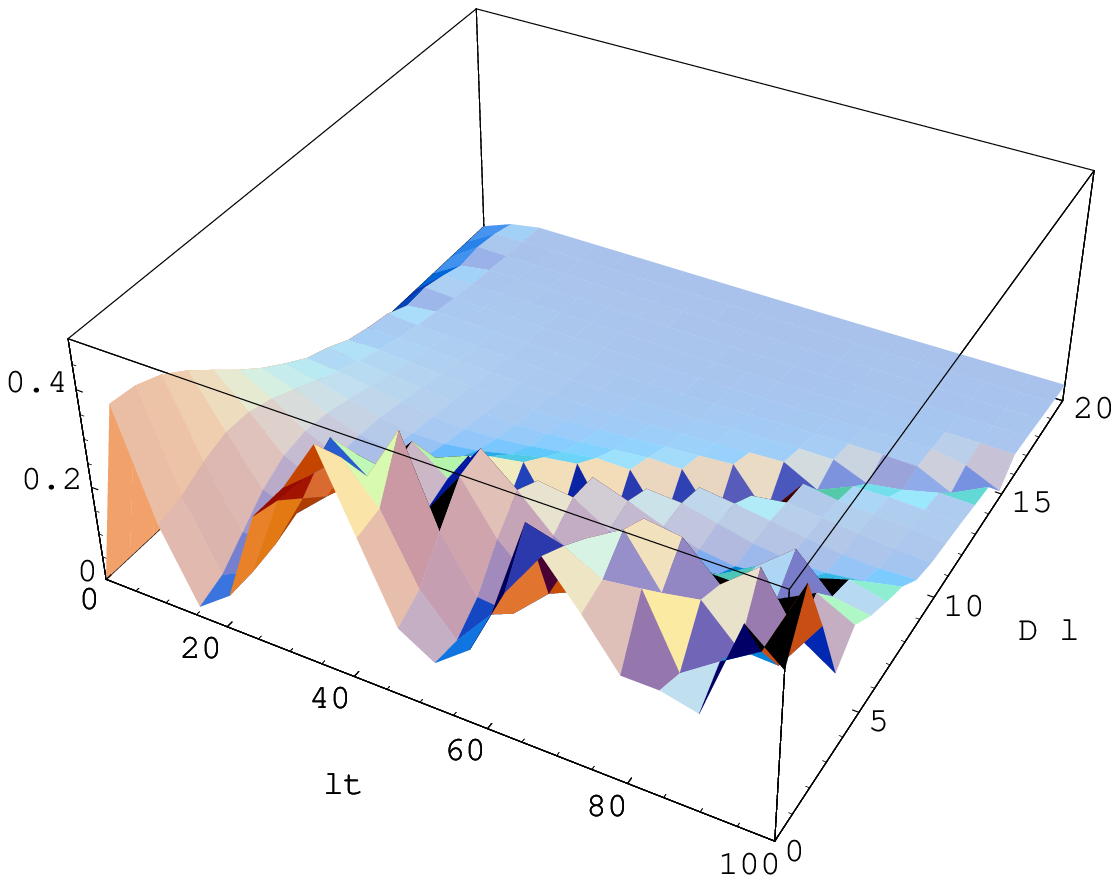} %
\includegraphics[width=10cm,height=6cm]{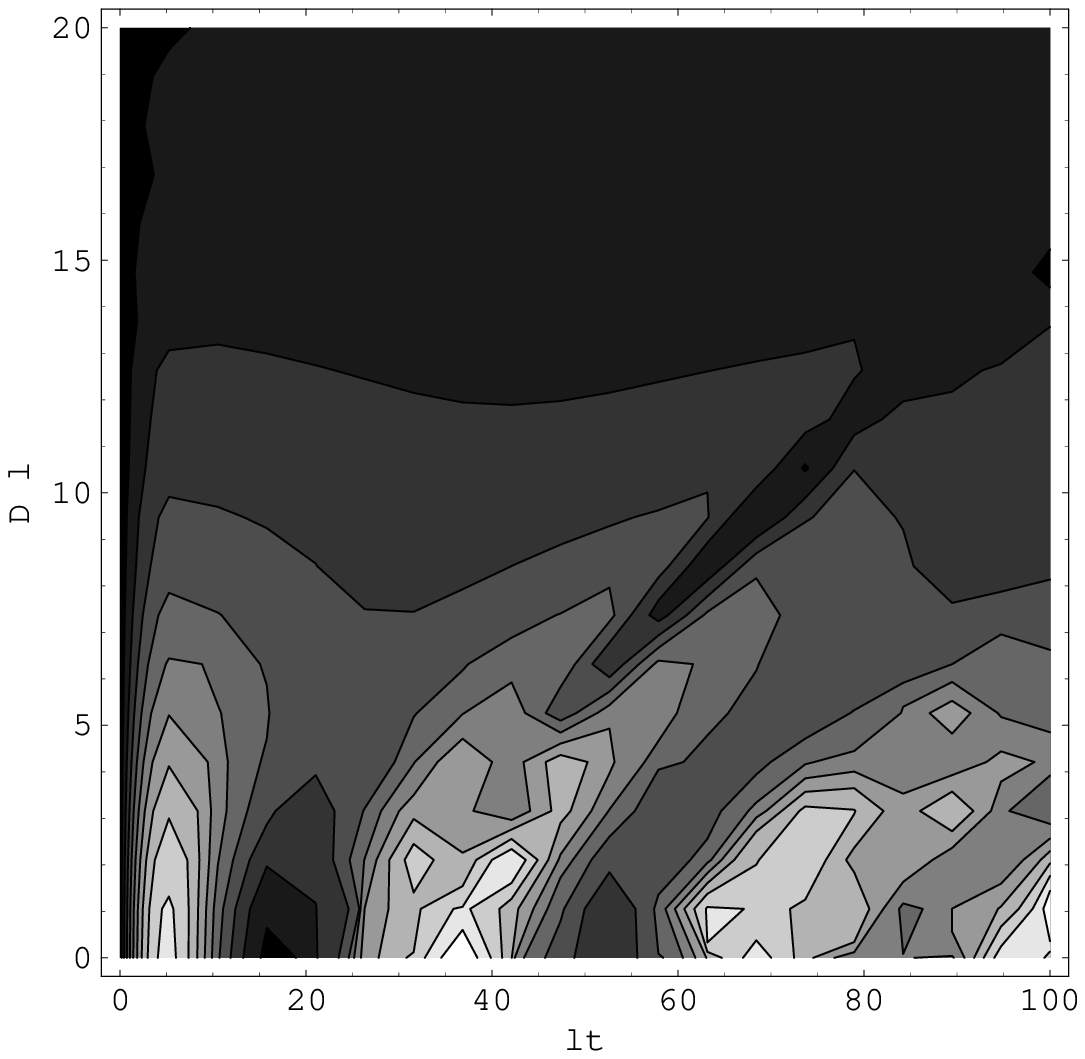} %
\includegraphics[width=11cm,height=6cm]{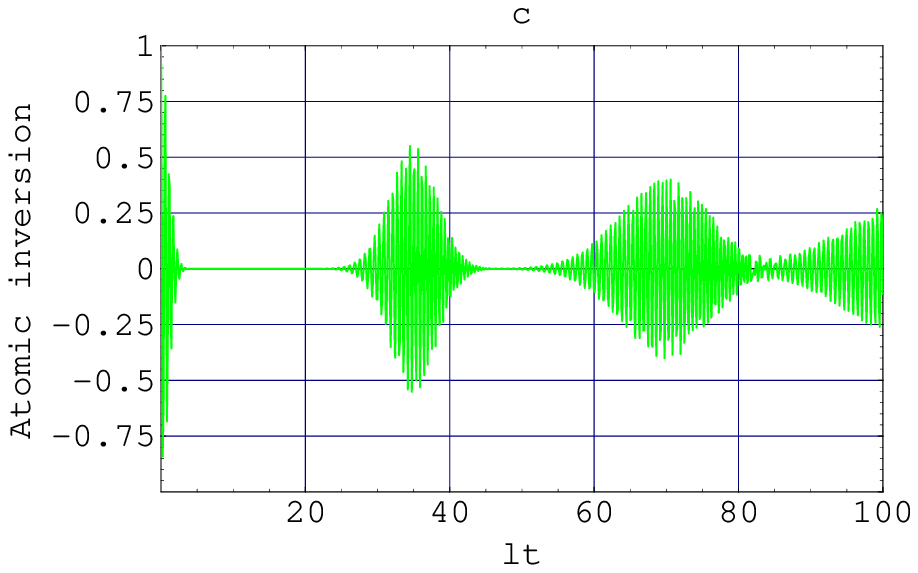}
\end{center}
\caption{(a,b) Plot of idempotency defect as a function of the dimensionless
scaled time $\protect\lambda t$ $(\protect\lambda =\protect\sqrt{e^{2}%
\protect\omega /(\hbar C_{F}})$ and the detuning parameter $\Delta /\protect%
\lambda $ and (c) the atomic inversion as a function of the scaled time $%
\protect\lambda t$. In the contour plot, disentanglement is shown in the
severe shading areas.}
\end{figure}

Supplemental to the analytical solution presented in the above
section, here we discuss the results obtained numerically and
interesting situations occurring for different values of the
detuning and phase-damped cavity parameters. We consider the
experimental parameters, described above, which are accessible using
the present day technology as, $C_{J}\sim 10^{-15}$F$,$ $\omega
\simeq 10^{10}$Hz, $C_{F}\sim 10^{-11}$F, $K_{B}T<<E_{J}$ $\sim
\hslash \omega <<E_{c}$ and the initial state of the filed as
coherent state. In order to analyze the effects resulting from
variation in the detuning or phase damped cavity we consider the
idempotency defect as a function of the scaled time $\lambda t$ and
$\Delta /\lambda $ ($\gamma /\lambda )$ shown in figures 2 and 3. We
have fixed the mean photon number of the coherent field as
$\bar{n}=10.$ As can be seen from figure 1, $\mathcal{E}_{t}^{(J)}$
smoothly diminishes with increasing the detuning parameter. For
further increasing of the detuning the impurity of the state of the
Cooper pair box system is rapidly growing and
$\mathcal{E}_{t}^{(J)}$ disappears completely. For the case when we
take $\gamma =\Delta =0.0$, we get almost zero values for the
idempotency defect only at $t=0,$ which means that a pure state will
not be reached at any time except at the initial stage of the
interaction time (see figure 1). To apprehend the essential features
of detuning effects on coherence loss, we presented in figure 1b the
contour plot of the concurrence, where complete separable states are
shown in the severe shading areas.

In Figure 2c we show the time evolution of the atomic inversion.
Apparently, it is easy to observe the existence of collapse and
revival of Rabi oscillations of the atomic inversion and the first
maximum of the idempotency defect is achieved in the collapse time,
while at one-half of the revival time, the idempotency defect
reaches its local minimum. Also, it is noticed that in absence of
both detuning and phase-damping, a gradual decrease in the
amplitudes of the Rabi oscillations is shown.

On the other hand, the decoherence introduces irreversibility into
the Junction dynamics and also on the global system. During the
repeated periods of maximum and minimum idempotency defect, the
states of the Junction and field lose and gain coherence, but given
the continuous amplitude decreasing of coherent states, the
coherence recovered by the Junction is never that which was lost. We
may refer here to the work given in reference \cite{ram07} where
engineering maximally entangled states has been discussed for
different systems. Of course, larger the value of $\gamma $, the
more rapid is this phenomenon in the sense of the idempotency
defects being close to one (the purity loss of the junction state is
complete). In particular, for the limiting case of large $\gamma $
($\gamma =0.1\lambda )$, the idempotency defect blows up from zero
and rapidly saturates i.e. as time goes on a long-lived coherence
loss is observed (see figure 2 (top)).
\begin{figure}[tbph]
\begin{center}
\includegraphics[width=10cm,height=8cm]{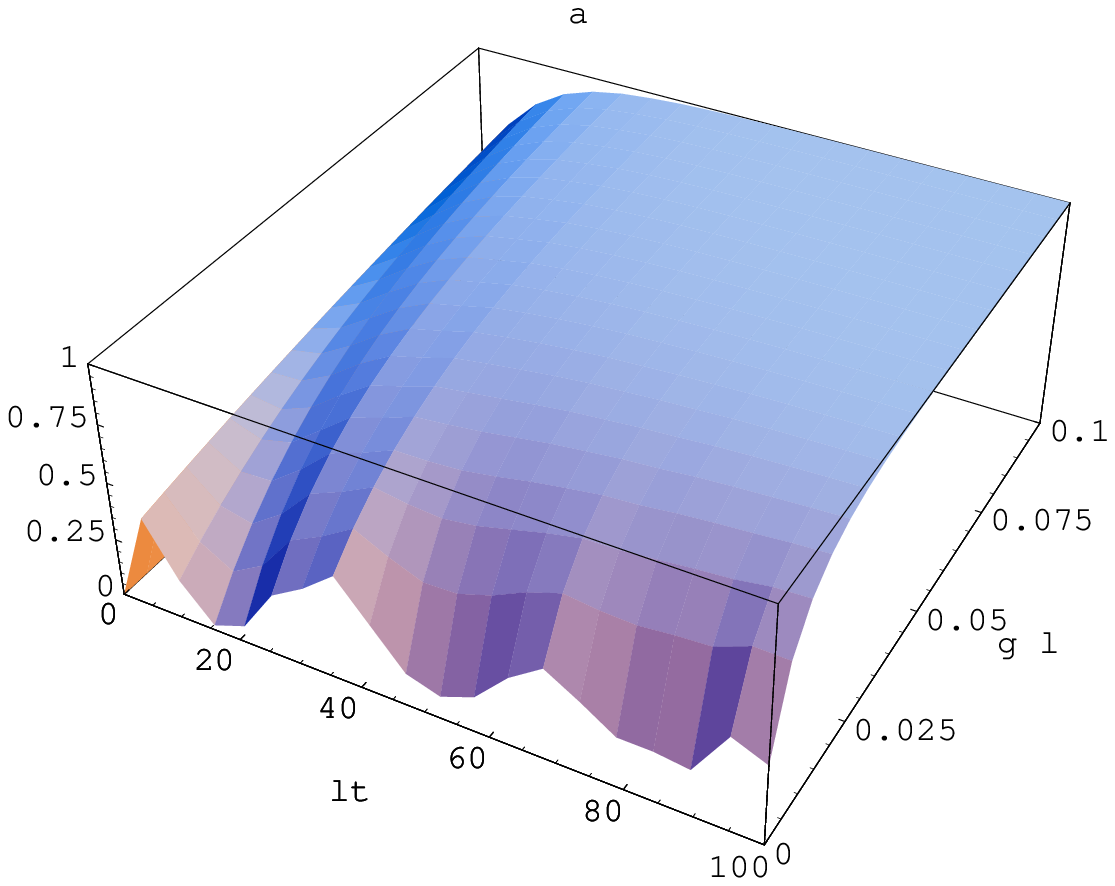} %
\includegraphics[width=10cm,height=6cm]{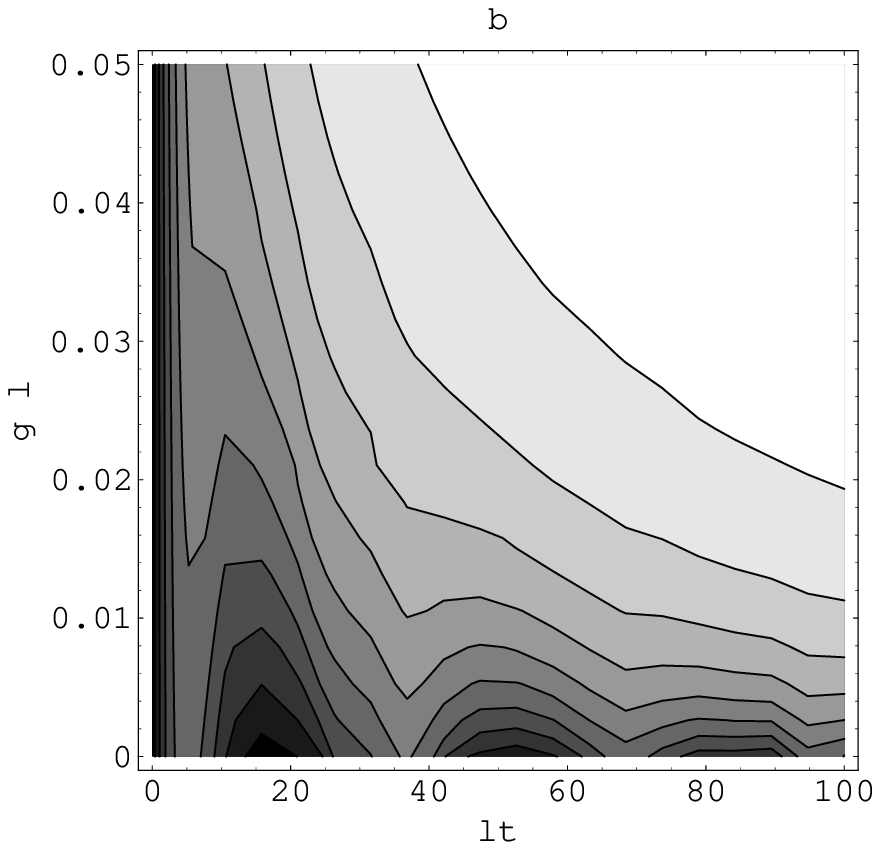}
\end{center}
\caption{Plot of idempotency defect as a function of the scaled time $%
\protect\lambda t$ and the decoherence parameter $\protect\gamma /\protect%
\lambda $. In the contour plot, a complete mixture is shown in the
non-shaded area. } \label{fig3}
\end{figure}
Even in a weak-damping cavity, the difference between consecutive local
maximum and minimum diminishes with time, since idempotency defect tends to
asymptotic values. Speaking specifically, it arrives at a maximum value
(about 1) at large values of the phase damping parameter, and then remains
nearly invariant regardless of the increase of time or $\gamma $, while the
idempotency defect always remains vanishing at $\lambda t=0$.

We can gain further physical insight into the dynamical effect of
the phase damping by considering the general case (mixed state
entanglement). To measure the degree of entanglement for mixed
states of bipartite systems composed by two-level subsystems, one
needs to consider a commonly used measure such as the concurrence
\cite{min05} which has been proven to be a reasonable entanglement
measure or negativity \cite{vid02}. Analysis of the entanglement
decay rates under decoherence for different models of the
interaction between systems of arbitrary dimensions with the
environment has been presented \cite{car07}. For the density matrix
$\hat{\rho}(t),$ which represents the state of a bipartite system,
concurrence is defined as
\begin{equation}
C(\hat{\rho})=\max \{0,\Re _{1}-\Re _{2}-\Re _{3}-\Re _{4}\},
\end{equation}%
where the $\Re _{i}$ are the non-negative eigenvalues, in decreasing
order ($\Re _{1}\geq \Re _{2}\geq \Re _{3}\geq \Re _{4}$), of the
Hermitian matrix $\widehat{\Upsilon }\equiv
\sqrt{\sqrt{\hat{\rho}}\widetilde{\rho }\sqrt{\hat{\rho}}}$ and
$\widetilde{\rho }=\left( \widehat{\sigma }_{y}\otimes
\widehat{\sigma }_{y}\right) \hat{\rho}^{\ast }\left(
\widehat{\sigma }_{y}\otimes \widehat{\sigma }_{y}\right)$. Here,
$\hat{\rho}^{\ast }$ represents the complex conjugate of the density
matrix $\hat{\rho}$ when it is expressed in a fixed basis and
$\widehat{\sigma }_{y}$ represents the Pauli matrix in the same
basis. The function $C(\hat{\rho})$ ranges from $0$ for a separable
state to $1$ for a maximum entanglement.
\begin{figure}[tbph]
\begin{center}
\includegraphics[width=10cm,height=8cm]{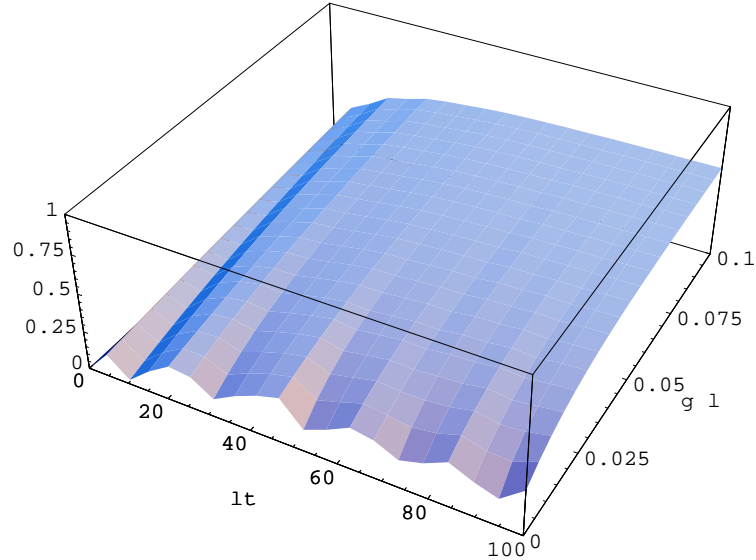}
\end{center}
\caption{Development of the concurrence $C(\widehat{\rho })$ as a
function of the scaled time $\lambda t$ and the decoherence
parameter $\gamma /\lambda $. The other parameters are the same as
figure 3. } \label{conc}
\end{figure}

In figure (\ref{conc}), we plot the numerically evaluated results
for the concurrence $C(\hat{\rho})$, as a function of the scaled
time $\lambda t$ and phase damping parameter $\gamma $ in units of
$\lambda $. As confirmed in figure (\ref{conc}), the asymptotic
value of the concurrence is obtained when the phase damping is
increased. Of course, there are some differences between the
concurrence and idempotency defect in the amplitudes but the general
behavior is the same i.e. comparing figures 3a and 4, one can find
that concurrence results in qualitative analogy with the results of
the idempotency defect. This may be thought to arise from the
asymptotic limits which have been observed in both figures 3 and 4
due to the phase damping. Although the entanglement, as witnessed by
the concurrence, is lower than maximal possible (about $1$), it has
a fixed value as the phase damping increased (long-lived
entanglement). We have confirmed the predictions of this phenomenon
using a systematic numerical analysis where a number of relevant
parameters has been varied. However, once the initial state setting
of the Cooper pair box is considered as $\rho
^{J}(0)=\cos^2(\theta)|e\rangle \langle e|+\sin^2(\theta)|g\rangle
\langle g|$, $(0<\theta<\pi/2$), this feature no longer exists and
entanglement vanishes in an asymptotic limit.

Obviously, the above novel phenomena are directly related to the
recent results of Ref. \cite{daj07}. With this at hand, one may
envision quantum computers using these long-lived entangled states
for quantum memory and for extended quantum information processing
\cite{haf05} where, superconducting single Cooper pair boxes using
superconducting single electron transistor fabricated on the same
chip as an electrometer has been presented in Ref. \cite{top01} and
the electronic control of a single-qubit achieved in a solid state
device has been demonstrated \cite{tsa01}. In these works, it has
been shown that, the general scalability of such a solid state
device will be a prerequisite for a practical quantum computer.
Also, it has been shown that \cite{haf05} only twice the resources
(qubits + elementary quantum gates in the decoherence free subspace)
are needed to realize up to 4 orders of magnitude more operations
before the quantum information is lost to the environment.

It has been predicted only recently that the one-body and two-body
responses to a noisy environment can follow surprisingly different
pathways to complete decoherence \cite{yu04,dod04}. The first
experimental work and impressive results in this new domain have
been reported in Ref. \cite{alm07}. They have devised an elegantly
clean way to check and to confirm the existence of so-called
entanglement sudden death, a two-body disentanglement that is novel
among known relaxation effects because it has no lifetime in any
usual sense, that is, entanglement terminates completely after a
finite interval, without a smoothly diminishing long-time tail
\cite{yu06,ebe07}.

\section{Conclusion}

In conclusion, we suggest that applying a microwave field to a
Cooper-pair box via the gate capacitance, a long-lived entanglement
can be realized i.e. the Cooper pair composite system is entangled
forever. In our work we have extended the exactly solvable model of
a single-Cooper-pair box model by taking into account the
decoherence effect on the purity loss and entanglement. Decoherence
is a very useful concept that has recently been widely investigated
and has turned out to be very prolific. It is intuitively related to
the loss of purity of a final state of the quantum system. However,
it is demonstrated that, it is not correct to think that a quantum
system, by increasing the decoherence, will suffer an increasing
loss of quantum coherence. It is worth stressing in this respect, an
appropriate choice of the system parameters, specifically, large
values of the decoherence parameter and initial state setting of the
Cooper pair box does give an interesting effect to the entanglement
process as a long-lived entanglement which may lead to unexpected
applications.

We are sure that our ground breaking work on the dynamics of quantum
entanglement in the Cooper pair box system exactly, will lead both
to understand the generic behaviors of theses systems by model
studies and to add more features to the theoretical models that can
provide a closer depiction of reality, captured in the near future
by higher precision experiments. A topic that remains open in almost
all decoherence discussions, however, is the preservation or
destruction of two-body quantum coherence when both bodies are
small. We are convinced that future experiments exploiting the
particular advantages of these models will reveal interesting new
phenomena and show many surprises.

{\bf Acknowledgement}

I would like to thank the referees for objective comments and for
bringing to my attention some new references. Also, I would like to
thank A. Buchleitner for very constructive suggestions and A.-S. F.
Obada and F. Saif for critical reading of the manuscript.

\[
\]
\textbf{References}
\[
\]

\end{document}